\documentclass{ieeeaccess}
\usepackage{cite}
\usepackage{amsmath,amssymb,amsfonts}
\usepackage{algorithmic}
\usepackage{graphicx}
\usepackage{subcaption}
\usepackage{varwidth}
\usepackage{caption}
\usepackage{dblfloatfix} 
\usepackage{textcomp}
\usepackage{xcolor}
\usepackage{braket}
\usepackage{soul}
\usepackage[urlcolor=blue]{hyperref}
\def\BibTeX{{\rm B\kern-.05em{\sc i\kern-.025em b}\kern-.08em
    T\kern-.1667em\lower.7ex\hbox{E}\kern-.125emX}}
\begin{document}
\history{Date of publication xxxx 00, 0000, date of current version xxxx 00, 0000.}
\doi{10.1109/TQE.2020.DOI}

\title{A Quantum Circuit Obfuscation Methodology for Security and Privacy}

\author{
\uppercase{Aakarshitha Suresh}\authorrefmark{1}, \IEEEmembership{Student Member, IEEE},
\uppercase{Abdullah Ash- Saki}\authorrefmark{2}, \IEEEmembership{Student Member, IEEE},
\uppercase{Mahabubul Alam}\authorrefmark{3},     \IEEEmembership{Student Member, IEEE},
\uppercase{Rasit O. Topaloglu}\authorrefmark{4}, \IEEEmembership{Senior Member, IEEE},
and \uppercase{Swaroop Ghosh}.\authorrefmark{5},
\IEEEmembership{Senior Member, IEEE}}
\address[1]{Pennsylvania State University, University Park, PA 16802 USA (email: ams9647@psu.edu)}
\address[2]{Pennsylvania State University, University Park, PA 16802 USA (email: axs1251@psu.edu)}
\address[3]{Pennsylvania State University, University Park, PA 16802 USA (email: mxa890@psu.edu)}
\address[4]{IBM, Hopewell Junction, NY 12533, USA (email:rasit@us.ibm.com)}
\address[5]{Pennsylvania State University, University Park, PA 16802 USA (email: szg212@psu.edu)}
\tfootnote{The work is supported by 
National Science Foundation (NSF) (CNS- 1722557, CCF-1718474, OIA-2040667, DGE-1723687 and DGE-1821766) and seed grants from Penn State Institute for Computational and Data Sciences (ICDS) and Huck Institute of the Life Sciences.}

\mark
{ AAuthor: Aakarshitha Suresh (email: ams9647@psu.edu).}
\begin{abstract}
Optimization of quantum circuits using an efficient compiler is key to its success for Noisy Intermediate Scale Quantum (NISQ) computers. Several $3^{rd}$ party compilers are evolving to offer improved circuit depth and faster compilation time and better depth/gate count for large quantum circuits. 
These $3^{rd}$ parties, or just a certain release of an otherwise trustworthy compiler, may possibly be untrusted and this could lead to an adversary to Reverse Engineer (RE) the quantum circuit for extracting sensitive aspects e.g., circuit topology, program, and its properties. In this paper, we propose obfuscation of quantum circuits to hide the functionality. Quantum circuits have inherent margin between correct and incorrect outputs. Therefore, obfuscation (i.e., corruption of functionality) by inserting dummy gates is nontrivial. We insert dummy SWAP gates one at a time for maximum corruption of functionality before sending the quantum circuit to an untrusted compiler. If an untrusted party clones the design, they  get incorrect functionality. The designer removes the dummy SWAP gate post-compilation to restore the correct functionality. Compared to a classical counterpart, the quantum chip does not reveal the circuit functionality. Therefore, an adversary cannot guess the SWAP gate and location or validate using an oracle model. Evaluation of realistic quantum circuit with or without SWAP insertion is impossible in classical computers. Therefore, we propose a metric-based SWAP gate insertion process. The objective of the metric is to ensure maximum corruption of functionality measured using Total Variation Distance (TVD). The proposed approach is validated using IBM default noisy simulation model. 
Our metric-based approach predicts the SWAP position to achieve TVD of up to $50\%$, and performs $7.5\%$ better than average TVD, and performs within $12.3\%$ of the best obtainable TVD for the benchmarks. We obtain an overhead of less than $5\%$ for the number of gates and circuit depth after addition of SWAP gates. 

\end{abstract}


\titlepgskip=-15pt

\maketitle
\section{Introduction}\label{intro}
\begin{figure*} [t] 
 \begin{center}
    \includegraphics[width=0.9\textwidth]{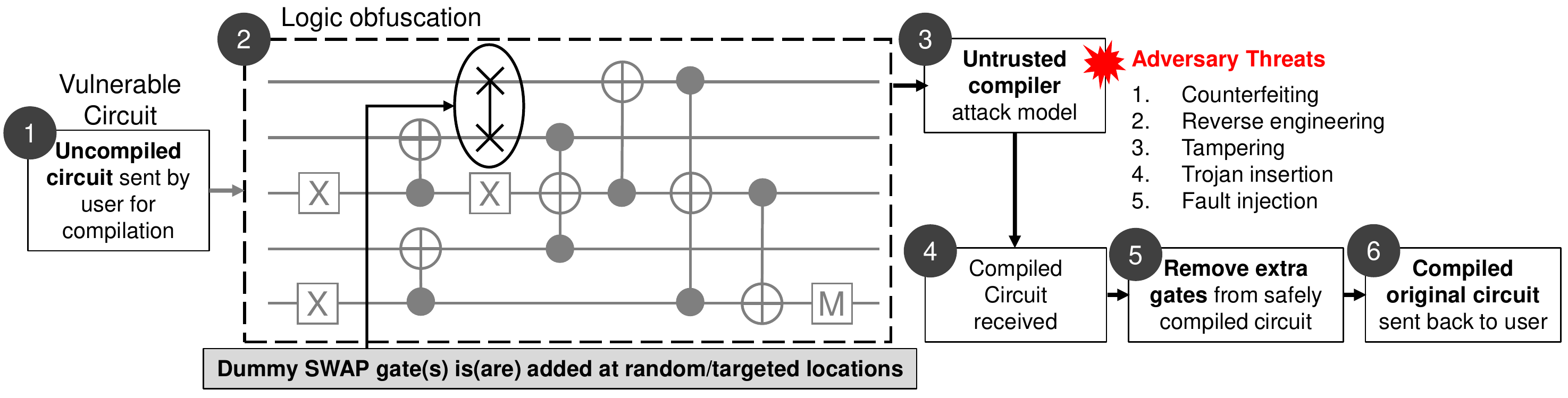}
 \end{center}
 \caption{Attack model proposed in this paper. The quantum circuit is sent by the user to the untrusted compiler, where the adversary can steal the IP or RE the circuit. Logic obfuscation is proposed as countermeasure.} \label{fig:atkmodel}
\end{figure*}

\PARstart{Q}{uantum Computing} can provide exponential speed-up over classical counterparts to solve certain class of combinatorial problems e.g., data analytics, material discovery, and drug synthesis. 
The computing power of quantum computers is growing due to rapidly evolving noise mitigation techniques \cite{NISQEraQC, VQGOpt}, ever-increasing number of qubits, and improving error rates and decoherence times \cite{LOQP}. Powerful gate-based universal quantum computers has potential to solve societal and science problems that are deemed hard by classical computers as also hinted by the Google experiment \cite{QSpsp}. Hybrid classical-quantum computing using shallow depth variational algorithms e.g., Quantum Approximate Optimization Algorithm (QAOA) and Variational Quantum Eigensolver (VQE) \cite{QAOA, HEVQE} has been explored to compute approximate solutions in presence of noise. Even if these circuits are shallow (to be meet the coherence time specification), the complexity can be very high depending on the problem and the number of required qubits. 
The problem-specific parametric quantum circuits designed using variational algorithms e.g., QAOA to solve certain problems embed the topology of the problem (an asset). For applications e.g., power grid (or other critical infrastructure) optimization, the client would like to keep the problem information confidential. Non-parametric circuits e.g., Quantum Fourier Transform (QFT) that are optimized for higher success probability with lower gate count and depth (to meet certain hardware coupling constraint) also contain Intellectual Property (IP). 
These IPs may not present risk for small scale quantum circuits that can be compiled on trusted vendors e.g., IBM and Rigetti. Considering the scaling trend of current Noisy Intermediate Scale Quantum (NISQ) computers, hardware architectures with more than 200-1000 qubits are predicted to become a reality by 2023 \cite{NSQC,tds}. 
\textbf{Motivation:} Success of quantum circuits on NISQ computers heavily relies on the quality of the optimization. Poorly optimized circuits even though functionally identical to an optimized circuit, will produce random results. This aspect is different from semiconductor circuit optimization which mostly focus on improving area, power and speed. 
Several $3^{rd}$ party compilers \cite{g1,g2} claim efficient optimization of complex circuits. These trends lead to dependence on untrusted $3^{rd}$ party compilers for improved circuit depth and faster compilation time even for large, complex and dense quantum circuits. This is primarily due to inability of trusted compilers to converge for large-scale quantum circuits with higher packing density \cite{CFQAOA}. An untrusted compiler can steal sensitive IP and problem properties.
Fig. \ref{fig:atkmodel} shows a simplified design flow to solve an optimization problem using quantum computing. Obfuscation is necessary to protect the IP while ensuring the correct compilation and functionality. Conventional obfuscation techniques e.g., hiding gate functionality is ineffective since that will hinder the optimization process. For example, if a CNOT gate is camouflaged as a Toffoli gate, it will lead to incorrect compilation.
This paper proposes a logic obfuscation approach to deal with the aforementioned threats. \textit{To the best of our knowledge, this is the first effort to identify a new security and privacy threat space in the quantum circuit compilation and develop countermeasures.}
\textbf{Proposed idea:} Although obfuscation can protect the IP, probabilistic nature of computation and inherent margin between correct/incorrect outputs makes functional corruption challenging even with addition of fake gates. For example, if the probability of 0 and 1 measurement of a single qubit circuit is 0.8 and 0.2, respectively (i.e., a margin of 0.6) then the obfuscation needs to overcome this margin to corrupt the functionality. Obfuscation can also degrade the circuit quality due to poor optimization. 
The proposed obfuscation technique inserts additional gate operations (e.g., SWAP gates) which are not part of the original circuit. We refer to these additional gate operations as \emph{dummy gates}. The circuit designer inserts these dummy gates to the original circuit using our proposed method before sending it to the $3^{rd}$ party compiler. The objective is to hide the true functionality of the circuit from the untrusted compiler. The adversary needs to identify and remove the dummy gates from an obfuscated circuit to extract the original circuit. This is a computationally hard problem since any gate can be a potential dummy gate. Furthermore, there is a lack of oracle model (the quantum chip implements functionality using microwave/laser pulses that are not publicly accessible) to validate the adversarial guess. Therefore, adversarial effort to RE the obfuscated design is high. Any attempt to reuse the circuit without removing the dummy gates will result in corrupted or severely degraded performance. SWAP gates are chosen as dummy gates due to ease of removal post-compilation by the designer and functional corruption by exchanging the states between two qubits.
To further illustrate the idea, we have executed the circuit shown (Fig. \ref{fig:atkmodel})
on default noisy simulation model of FakeValencia with and without the dummy SWAP gate. The outputs are obtained for 8 possible inputs. For example, we observe \{``00000":10212, ``10000":7889, ``10100":5084, ``11000":10057, ``11100":5227, ``00100":7186, ``01000":46956, ``01100":7389\} in (Fig. \ref{fig:ckt}) for 123 counter circuit as correct outcome over 100000 shots. Here the first and second numbers separated by colon are measured output bits and observation frequency, respectively.
However, we observe \{``00000":14672, ``10000":3969, ``10100":6662, ``11000":2319, ``11100":1646, ``00100":53046, ``01000":6458, ``01100":11228\} after inserting dummy gate at position $6$. The obfuscated output distribution is quite different from the original circuit output distribution which indicates obfuscation of functionality. We use Total Variation Distance (TVD) as a measure of the degradation or the difference between obfuscated output and original output. TVD is defined as $\sum_i (|x_{orig, i} - x_{obfus, i}| \div shots)$. Here, $x_{i}$ is the count of $i^{th}$ element of a distribution.
One of the challenges with this technique is the lack of knowledge of the correct output state for a realistic optimization problem that is solved using the quantum circuit. Therefore, the designer cannot evaluate the effectiveness of his obfuscation using simulations. We propose several metrics to identify the right location where the dummy gate should be inserted to corrupt the output with high likelihood. Such a metric is identified from the insights developed from an exhaustive analysis of small quantum circuit benchmarks for various dummy gate insertion points evaluated using simulations.   
\textbf{Main contributions}: We, (a) propose novel threat model, (b) propose two qubit SWAP gate as a dummy gate-based obfuscation technique, (c) perform exhaustive experiments on a set of benchmarks, (d) propose features e.g., depth of circuit for each position and number of control qubits as metrics to select the SWAP gate location, (e) present extensive analysis and validation of the proposed metric-guided insertion of SWAP gates. 
\textbf{Paper organization:} We review the background material in Section \ref{basics}. Sections \ref{threats} and \ref{proposal} present the threat model and the proposed obfuscation procedure, respectively. Section \ref{results} focuses on the simulation results and analysis. The discussions are presented in Section \ref{disc} and conclusions are drawn in Section \ref{conc}.
\section{Background} \label{basics}
\subsection{Quantum Computing Preliminaries}
{\bf{Qubits:}}
Qubit is analogous to classical bits 0 and 1 with a key difference: qubit can be in a superposition state i.e., a combination of 0 and 1 at the same time. Qubit state is expressed with a \textit{ket} ($\ket{.}$) notation. A qubit state $\ket{\psi}$ is described as $\ket{\psi} = a \ket{0} + b \ket{1}$. Here, $\ket{0}$ and $\ket{1}$ are known as computational \textit{basis states} and $a$ and $b$ are complex numbers such that $|a|^2 + |b|^2 = 1$.
{\bf{Quantum gates:}}
Quantum gates are the operations that modulate the state of qubits and thus, perform computations. Quantum gates are represented by $2^n \times 2^n$ unitary matrices (n = number of qubits) and can work on a single qubit (e.g., X (NOT) gate) or on multiple qubits (e.g., 2-qubit CNOT gates).

{\bf{Basis gates and coupling constraints:}}
A practical quantum computer normally supports a limited number of single and multi-qubit gates known as \textit{basis gates} or native gates of the hardware. For instance, IBM quantum computers have the following basis gates: u1, u2, u3, id (single-qubit), and CNOT (two-qubit). However, the quantum circuit may contain high-level gates that are not native to the hardware e.g., the Toffoli gate in Fig. \ref{fig:atkmodel} is not native to the IBM quantum computers. Therefore, the gates in a quantum circuit are \textit{decomposed} into the basis gates before execution. Besides, the two-qubit operation (CNOT) is only permitted between the connected qubits.
These limitations in two-qubit operations in any target hardware are also known as \textit{coupling constraints}. 

{\bf{Compilation:}} Quantum circuit compilers e.g., Qiskit \cite{cross2018ibm} perform necessary modifications (e.g., insert SWAP gates) to the input circuits to meet coupling constraints of the hardware. Besides, compilers offer higher-level circuit optimization through single/multi-qubit gate cancellation, rotation merging and gate-reordering \cite{nam2018automated}. Qiskit supports barrier between circuit partitions to limit these additional optimizations across circuit partitions \cite{cross2018ibm}.

\subsection{Total Variation Distance (TVD)} 
TVD is a widely used metric to measure the difference between two quantum states \cite{Sarovar2020detectingcrosstalk, ash2020experimental}. We use TVD as a measure of the distance between the desired output state of the original circuit and the corrupted output state of the obfuscated circuit. A higher TVD indicates functional corruption of the circuit. Therefore, a higher TVD is desired for stronger obfuscation.

\subsection{Related Work} 
The works in\cite{cuiarticle42}- \cite{limaye44} assume that adversary will insert trojans in the reversible circuit before fabrication and send it back to the design companies. However, this attack model is not applicable to quantum circuits since basis gates are realized using microwave/laser pulse. The quantum circuit is never physically fabricated in gate-based model of quantum computing even though a quantum circuit is reversible. 
Binary data and test pattern-based detection assumptions made in the work does not apply to quantum circuits. 

Another work \cite{saeed149} assumes untrusted foundry that can locate ancillary and garbage lines in reversible circuit and can extract the circuit functionality. Dummy ancillary and garbage lines are added to the circuit which increases the ancillary and garbage lines post-synthesis. The attacker can identify only ancillary and garbage lines added post-synthesis, not the pre-synthesis. To reduce the overhead, reversible gates are added to the circuit judiciously to remove the ``telltale'' signs post-synthesis while keeping the logical functionality intact. The authors mentioned these approaches are only applicable for oracle-type or pure Boolean logic-based quantum circuits and not for quantum computing. Our proposed approach is more generally applicable for any quantum circuits including circuits with and without ancillary and garbage lines whereas \cite{saeed149} critically depends on the presence of the ancillary and garbage lines. Again, the work in \cite{saeed149} assumes the foundry and fabrication process of the reversible circuit is untrusted which is not directly applicable for gate-based quantum computing for reasons previously stated. In our work, we assume the compiler is untrusted.

Another work \cite{acharya148} assumes that malicious adversary in quantum cloud will report incorrect qubit quality to force erroneous computation. 
Our adversarial model considers the compiler to be untrusted whereas the model in \cite{acharya148} considers quantum cloud to be malicious. Therefore, we are addressing vulnerability in a different layer in the quantum computing stack.



\section{Threat Model} \label{threats}
\textbf{Motivation:} Quantum circuits can be lucrative targets of stealth for making profit if they are reused in multiple applications e.g., quantum Machine Learning (ML) circuits like classifiers. Problem-specific circuits optimized to solve the problem at scale can also be considered as IP. 
Furthermore, leaking of high-level information e.g., type of algorithm used in the circuit and the problem that is being solved can provide undue financial/political advantage to the adversary and compromise national security (depending on the problem being solved by the quantum circuit).

\textbf{Threats involved and feasibility}:
One of the important aspects of quantum circuit compilation is to optimize the circuit for improved circuit depth and reduced gate count. Several $3^{rd}$ party compilers are evolving that offer optimization at faster compilation time even for large quantum circuits \cite{g1,g2}. Following factors will motivate the quantum circuit designers to avail the untrusted $3^{rd}$ party compilation services, (a) success of quantum circuit, since optimized circuit is essential to obtain meaningful results from NISQ computers. Poorly optimized circuit even though functionally identical to optimized circuit will produce random outputs; (b) lack of trusted compilers that have caught up with the latest advancements in optimization; (c) availability of efficient but untrusted $3^{rd}$ party compilers 
(\cite{multiprogQC,g1,g2}) that are being developed to optimize depth and gate count compared to trusted compilers.
These compilers can be hosted on either the local machines by the $3^{rd}$ party or on the cloud service providers \cite{amazon} to launch, (i) cloning/counterfeiting, where quantum circuit can be stolen or reproduced; and (ii) Reverse Engineering (RE), where the sensitive aspects of the quantum circuit could be extracted. 

\section{Proposed Obfuscation Procedure} \label{proposal}
In this section, we provide overview of the proposed obfuscation procedure. First we describe various features of the dummy SWAP locations using examples. Next we explain various approaches to combine them to distill effective metrics. An example circuit is also shown to illustrate the features and performance of the proposed metrics.

\subsection{Overview}
Our main aim is to introduce dummy gates in a quantum circuit before it is sent to the untrusted compiler. 
This insertion should be done strategically to obtain a good TVD from the original circuit.
The approach is to exhaustively analyze all possible dummy gate insertion positions, the corresponding TVD values and identify their features. 
Once the features for a particular SWAP location and corresponding TVD is understood, a guided heuristic can be developed to insert dummy gates for a fresh circuit to maximize corruption or obfuscation of the circuit. 
 
 For the exhaustive search based analysis, a quantum circuit is chosen (here circuits with only two and three qubits controlled CNOT gates and double controlled CNOT gates (CX, CCX) are presently considered), and two qubit SWAP gates are inserted at locations without altering the circuit depth.
 To do this, the circuit is first divided into \emph{slices}. Inside each slice, any two quantum gates operate on different set of qubits meaning the quantum gates can operate in parallel. The impact of long distance swap gates (between non-neighboring qubits) also needs to be included in the exhaustive search-based analysis. Hence, assuming $n$ free qubits in a slice, there are $\binom{n}{2}$ possible ways to insert the dummy 2-qubit SWAP into the slice if $n \geq 2$ (even on non-neighbouring qubits). The obfuscated circuits contain one dummy gate per experimental trial.
 To study appropriate positions for the dummy gates, we insert one dummy SWAP gate at each of the possible positions of the benchmark circuits 

from Revlib \cite{wille2008revlib} repository (Fig. \ref{fig:atkmodel}) and analyze its impact on the output in terms of TVD to create a location prediction metric. These metrics are then studied to validate their effectiveness.


{\bf{Average, Worst, and Best TVD:}} We refer to the highest and the lowest TVD observed for each benchmark during exhaustive simulation as the Best, and the Worst TVDs, respectively. We term the average value of all the TVDs during the exhaustive analysis as Average TVD.

\subsection{Proposed Heuristic}



\textbf{Step -- 1:} 
First step is to identify various features that can distinguish the dummy gate locations from each other in a given circuit in terms of TVD. Some examples of these features include number of control gates on the qubits and depth of the dummy SWAP gate from output (explained next). 
These features are used individually or in combination to determine the best metric (Table \ref{fig:met}) for guided selection of SWAP gate location for future unseen circuits. The features are explained below. Fig. \ref{fig:ckt} illustrates the corresponding values for an example benchmark.  


\begin{figure}[b]
    \centering
    \includegraphics[width=3.3in]{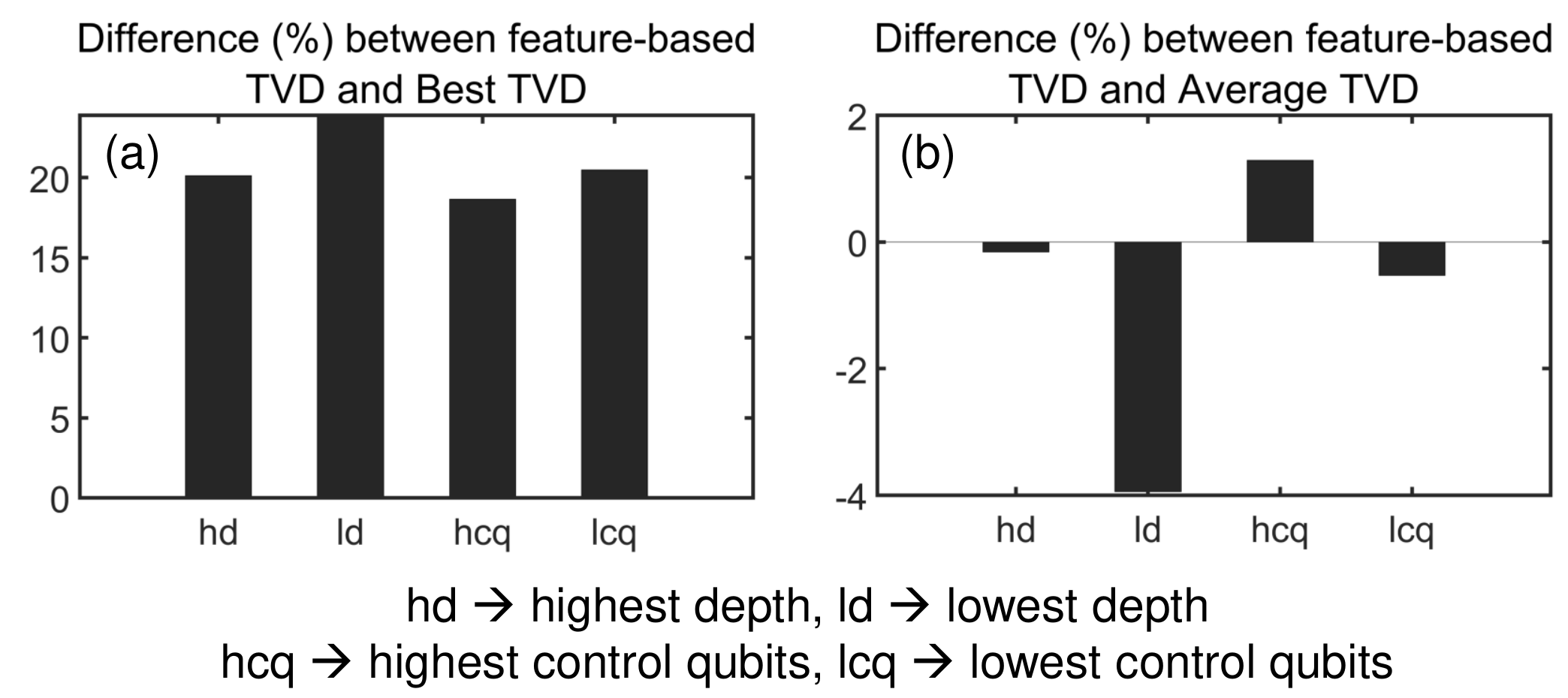}
    \caption{Difference (\%) between feature-based TVDs and (a) the Best TVD, and (b) the Average TVD.}
    \label{fig:pl2}
\end{figure}

\subsubsection{Depth of the dummy gate from primary outputs} 
\begin{figure*}
    \centering
    \includegraphics[width=7in]{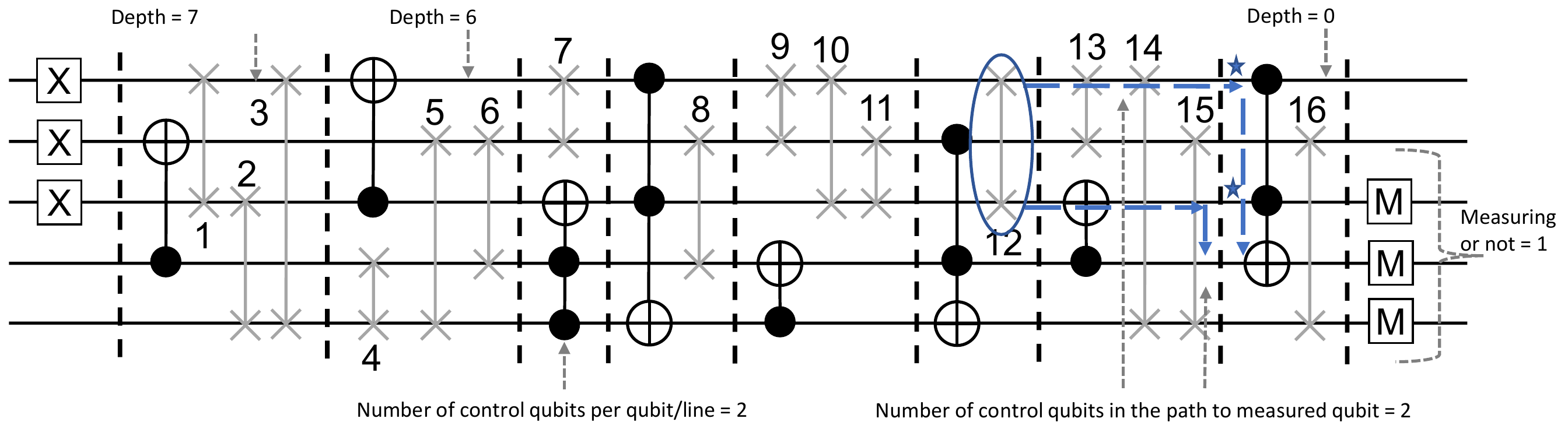}
    \caption{Circuit diagram of 123 counter with annotated features. For example, SWAP gate position 12 has depth = 2, number of control qubits = 5, measured or not = 1, constant qubits involved or not = 1, and number of control qubits in paths = 2, as shown. The path shown in blue, starts from each qubit of SWAP gate 12, and is continued in links by the control qubits, ending in target qubit that is measured. Doing this for both qubits in SWAP gate 12, we get 2 control qubits in the paths.}
    \label{fig:ckt}
\end{figure*}

We use this feature to calculate the number of slices present between the considered position and the output. It quantifies the influence of the SWAP gate on the output of the circuit. For example, the depths of SWAP positions 1, 2, and 4 are 7, 7, and 6 respectively (Fig.~\ref{fig:ckt}).

\subsubsection{Measuring qubit}
This feature checks if the two qubits associated with the dummy SWAP gate are being measured eventually or not.
If both of these qubits are measured, then it is likely that the SWAP gate will impact the output significantly. The impact reduces if only one of the qubits is measured. 
For example, SWAP--1 involves 1 measured qubit (Q2; qubits in the figure are indexed from Q0 to Q4 from the top), SWAP--2 involves 2 measured qubits (Q2 and Q4), and SWAP--7 involves 0 measured qubits.
\begin{figure*}
    \centering
    \includegraphics[width=7in]{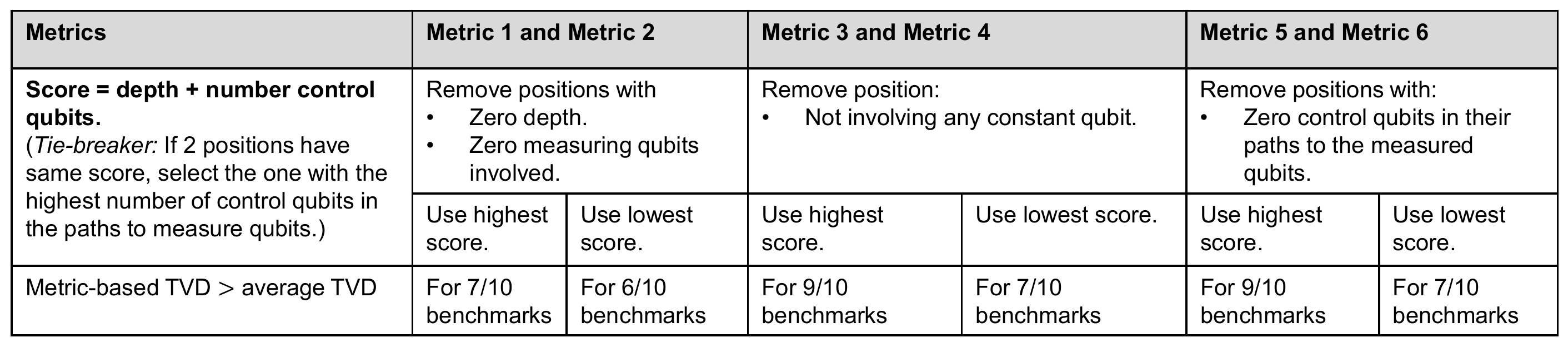}
    \captionof{table}{SWAP gate location selection metrics. }
    \label{fig:met}
\end{figure*}
\subsubsection{Number of times the qubit is used as control qubit} 
This feature counts the number of times the qubit involved in the dummy SWAP gate is used as control qubit in the circuit. 
This intuitively states that if this qubit is impacted, the other directly/indirectly controlled qubits will be affected as well.

\subsubsection{Number of control qubits in the paths}\label{subsubsec:num-ctrl}  This feature counts the number of control qubits that can be traced in the paths from each qubit involved in the dummy gate (source) till one of the measured output qubits is reached (destination). Multiple paths can be traced from the two qubits in a dummy gate to one of the destination measured qubit. The number of these control gates are added together for each dummy gate position. Consider SWAP--12 in Fig.~\ref{fig:ckt}. Two paths (one ``top'' and another ``bottom'') stemming from 2 qubits in the SWAP gate are shown in blue color. A path splits into two direction at a control qubit. One proceeds to the next time-step (if the next time-step has gates in it) and another takes a $90^\circ$ bend towards the target qubit. For SWAP--12, the ``top'' path splits into two. The split path proceeding to the next time-step has no gates in that time-step and hence, ignored. Another split path takes the $90^\circ$ bend and continues towards the target qubit, and in the process encounters another control qubit. Thus, the ``top'' path from SWAP--12 has 2 control qubits. Likewise, the `bottom'' encounters 1 control qubit during traversal. However, that control qubit is already accounted during the ``top'' path traversal, and hence, not counted again.


\subsubsection{Constant qubit:} Some circuits have constant values for some qubits e.g., 1 or 0. These constants affect the output especially, (i) for some input combinations; (ii) when the dummy gate involves one constant qubit. They get swapped and as a result, affect the other gates, eventually corrupting the output. Hence whether the dummy gate has a constant qubit involved is used as a feature.

\textbf{Step -- 2:}
After collecting all the features described above, we focus on using them to develop an effective metric. Initially, the dummy gate depth (or number of control qubit) feature is used individually to select the dummy gate location. However, these simple metrics produce multiple positions with the same feature values. The TVD of these positions are averaged and compared with the average and the best TVD of each of the benchmark circuit. Fig. \ref{fig:pl2} shows the average across all benchmarks for 2 features namely, depth and number of control qubits. It can be noted that each of these features exhibit strong correlation towards output TVD. Further, some features might maximize the TVD more than others. Since these features produce more than one positions, we use their combination to distill a single best SWAP location for each circuit. Fig.~\ref{fig:pl2} shows, if we consider a single feature, the difference from best TVD is relatively high ($\approx 20\%-25\%$, Fig.~\ref{fig:pl2}a) and it lags behind the average TVD in most cases (Fig.~\ref{fig:pl2}b).
Therefore, we consider multiple features together to refine our choice. This approach eliminates few positions and safely reduces the search space for each of the benchmarks. We aim to develop a metric that provides at least $5\%$ better TVD than average TVD and not worse than $15\%$ from the best TVD for most of the circuits.

Table~\ref{fig:met} summarizes various metrics used in this paper. The metrics are different heuristics guided by simulation results to filter out ineffective SWAP positions for obfuscation. 

\textbf{Metric-1 and Metric-2:} In the first pass, we remove SWAP positions with 0 depth and 0 measuring qubits involved. These cases are SWAP gates at primary output involving qubits that are not measured. 
The corresponding TVDs attest to the fact that such SWAP positions are ineffective for obfuscation and can be safely removed from consideration. After this pruning, we obtain a reduced set of SWAPs. We compute \emph{score} for each positions to select candidate SWAP location. Metric-1 picks the SWAP with highest score and Metric-2 picks the one with lowest score.

\textbf{Metric-3 and Metric-4:} In the second pass (i.e., with after SWAP list reduction employed in Metric-1/2), we remove SWAP(s) not involving any constant qubits to further sanitize the SWAP list. Metrics-3 and 4 are based on the highest and lowest scores, respectively.

\textbf{Metric-5 and Metric-6:} Finally, we remove SWAP locations involving 0 control gates in the path to a measured qubit (described in Section \ref{subsubsec:num-ctrl}). The potential impact of the SWAP on the output is minimum without any control operations in the path. Thus, such SWAP locations are not ideal for obfuscation. Metrics-5 and 6 are based on the highest and lowest scores, respectively. 

In denser circuits, higher depth means that the SWAP gate is inserted earlier in the circuit, and higher number of control qubits both add up to give highest score. Also note that, lowest score either indicates lower depth implying that the SWAP gate is located closer to the measured qubits and/or that the number of control qubits is less and can directly impact the output TVD significantly. We consider both highest and lowest scores in our metrics. However, we find that the highest score based metric (Metric 5) performs better on more number of circuits. Therefore, this can be the metric of choice.

\subsection{Case Study- 123 Counter circuit}

\begin{figure*}[t]
    \centering
    \includegraphics[width=0.8\linewidth]{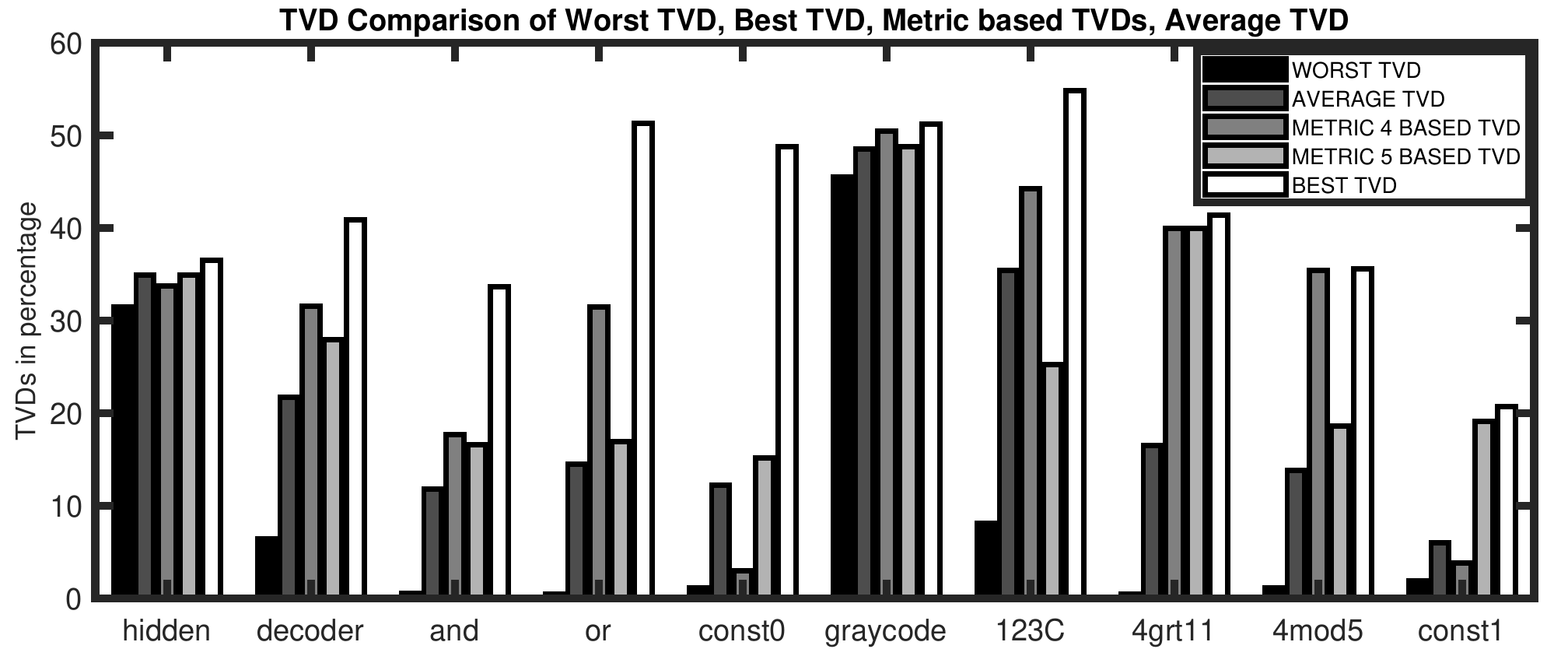}
    \caption{Plot for worst TVDs, best TVDs, metric-based TVDs, and average TVD for all ten circuits.}
    \label{fig:pl1}
\end{figure*}

The 123 counter circuit (Fig. \ref{fig:ckt}) is divided into $8$ slices between barriers. For each experiment, one SWAP gate at a time is inserted into one of the $16$ possible locations without overlapping with the other gates in the slices. After compilation and simulation for individual SWAP gate positions, it has been noted that the worst, best, and average TVD are $8.1\%$, $54.86\%$, and $35.45\%$, respectively.
As shown in Fig. \ref{fig:pl2}, the best performing feature based dummy gate selection is the number of control qubits feature which is only $18.66\%$ worse than the best TVD and $1.29\%$ better than the average TVD. 
However, this feature cannot be used in isolation since it provides multiple positions with same number of control qubits. To solve this issue various combinations of the features like depth, number of control qubits, constant or not, etc. (Table \ref{fig:met}) are used to shortlist the positions that can potentially provide best performance. 
Our initial metric employed (\textit{score = depth feature + number of control qubits feature}) and selected the highest score positions. This metric performed $24.7\%$ worse than best TVD and $5.29\%$ worse than the average TVD. Clearly, this simple metric did not outperform average TVD. Therefore, we combine few features and next prune the list of SWAP gate positions. Finally, we use a scoring function to select the best position from the list. In this example circuit, Metric $4$ outperformed average TVD by $8.76\%$  and under performed only $10.65\%$ than the best TVD. Metric $5$ under performs by $10.2\%$ and $27\%$ than the average and best TVD, respectively. Hence, metric $4$ is better choice for this example circuit. However, the final metric should be generic i.e., it should outperform the average TVD for most of the circuit and should be as close to best TVD as possible. Our analysis indicates that metric $5$ meets this requirement even though it performs poorly for this particular example circuit. 
\section{Experimental Results and Analysis} \label{results}
Note that, one can perform exhaustive simulation in order to find the best positions to insert the dummy gates. However, simulation is prohibitively expensive for quantum circuits with large number of qubits. Moreover, the application loses any perceived quantum advantage if it can be classically simulated. 
One of the prime objectives of this article is to devise a scalable approach for quantum circuit obfuscation that can provide obfuscation strength comparable to the exhaustive approach. In this Section, we show results and analysis on small quantum circuit benchmarks to demonstrate the benefit of the proposed heuristics. Before going into the details, we first provide a brief description of our experimental/simulation framework, and benchmark.

\begin{figure}[t]
    \centering
    \includegraphics[width=3.2in]{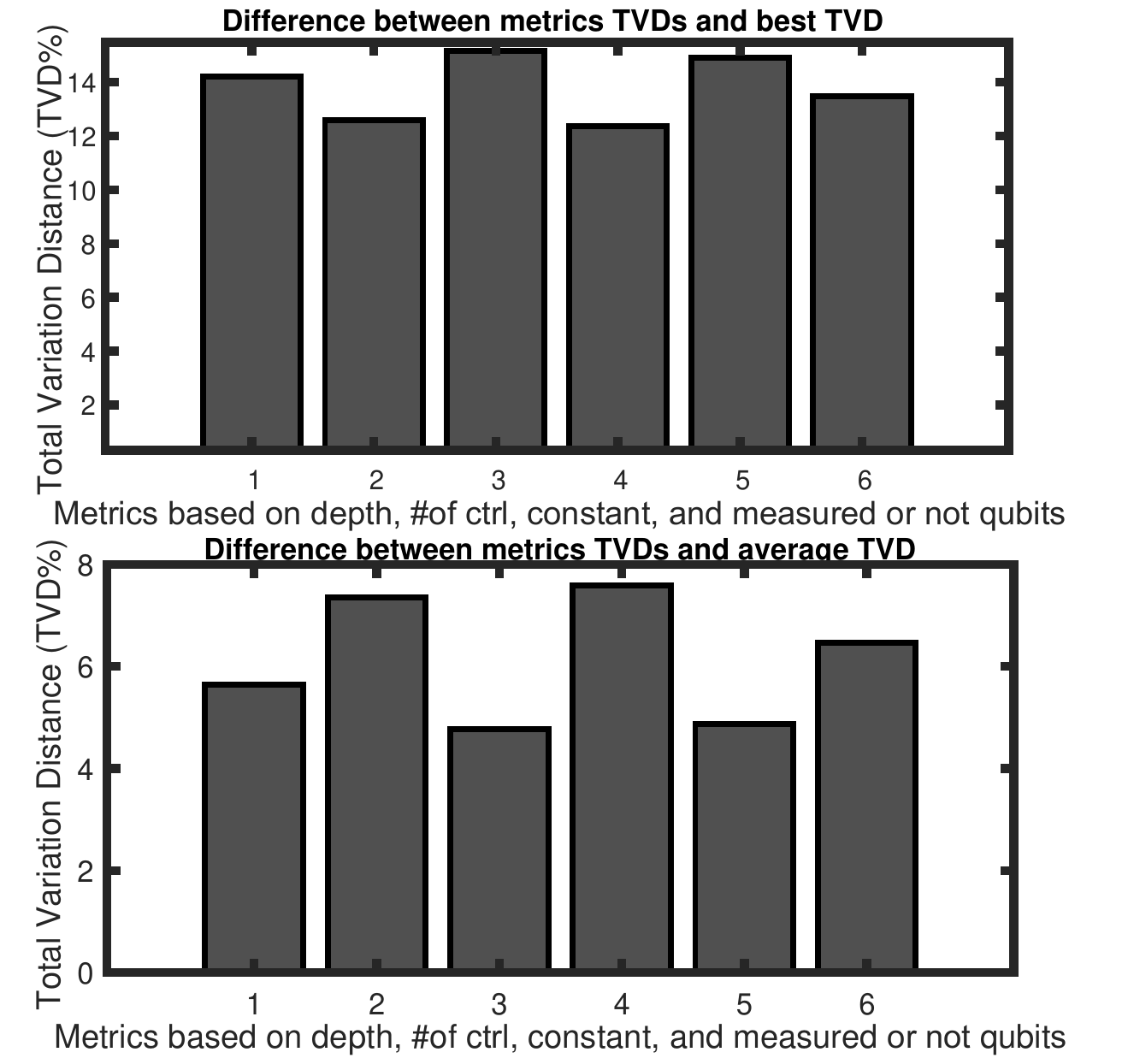}
    \caption{Difference (\%) between metrics-based TVDs and (a) the Best TVD, and (b) the Average TVD (Metrics 1, 2, 3, 4, 5, \& 6).}
    \label{fig:pl3}
\end{figure}

{\textbf{Evaluation Framework and Benchmarks:}} We use the open-source quantum software development kit from IBM (Qiskit) \cite{cross2018ibm} for our simulations. The software runs locally on an Intel Core i5-9300H CPU clocked at 2.40GHz machine. The default compiler backend in Qiskit is used for compilation.
A Python-based wrapper is built around Qiskit to accommodate the proposed obfuscation on the input circuits to the compiler backend. We use $10$ reversible circuit benchmarks from the RevLib repository \cite{wille2008revlib} which are commonly used in the contemporary works on quantum circuit compilation \cite{multiprogQC, NISQEraQC, ash2019qure}. We use the noisy hardware simulation framework available in Qiskit with the realistic noise values from ibmq\_valencia for all the simulations \cite{cross2018ibm}.
{\bf{Comparative Analysis:}} Fig. \ref{fig:pl1} shows the Worst, the Best, the Average TVDs of 10 reversible circuit benchmarks alongside the TVDs observed with two of our proposed heuristics - Metric 4 and Metric 5. Note that, for most of the benchmarks, both these heuristics provided higher TVD's than the Average TVD. Metric 4 and 5 provided $7.6\%$ and $4.9\%$ higher TVD, respectively than the Average TVD over all the benchmarks. They under performed $12.37\%$ and $14.91\%$ on average, respectively from the Best TVDs.
We show the average distance between the Best TVD and the metric-based TVDs across all 10 benchmarks in Fig. \ref{fig:pl3}(a). Note that, a higher value indicates a lesser performance. Metric 2 and 4 performed better than the rest with distances of 12.60\% and 12.37\%, respectively. Metric 3 performed poorly with a distance of 15.18\%. In Fig. \ref{fig:pl3}(b), we show the mean distance between the Average TVD and the metric-based TVDs. Note that, a higher distance value indicates a better performance in this case. Again, Metric 2 and 4 performed better than the rest with average distances of 7.35\% and 7.58\% each. Metric 3 performed poorly with average distance of 4.77\%. The results indicate that Metric 2 and 4 can be better choices for strong obfuscation. Metric 1 also performed poorly with a distance of $14.226\%$ from the best TVD and $5.65\%$ from the Average TVD. Metric 6 performed better than Metric 1, but worse than metrics 2, 4 or 5 on average.
Ideally, we expect to get TVDs greater than the Average TVDs and closer to the Best TVDs from the heuristic solutions to signify that the heuristic performs better than a random approach. However, none of the metrics provided consistent performance to that front in our simulations. For instance, Metric 4 provided better than the Average TVDs for 7 out of the 10 benchmarks whereas Metric 5 provided better TVD in 9 out of 10. Therefore, it is evident that a single metric may not provide decent performance over all the benchmarks. One potential direction is to incorporate multiple metrics in the obfuscation procedure.
For instance, we can apply both Metric 4 and metric 5 on the same circuit one after another. Unless these metrics fail to corrupt the outputs sufficiently, such an approach can provide consistent performance at the expense of extra dummy gates.
\begin{figure}[b]
    \centering
    \includegraphics[width=3.3in]{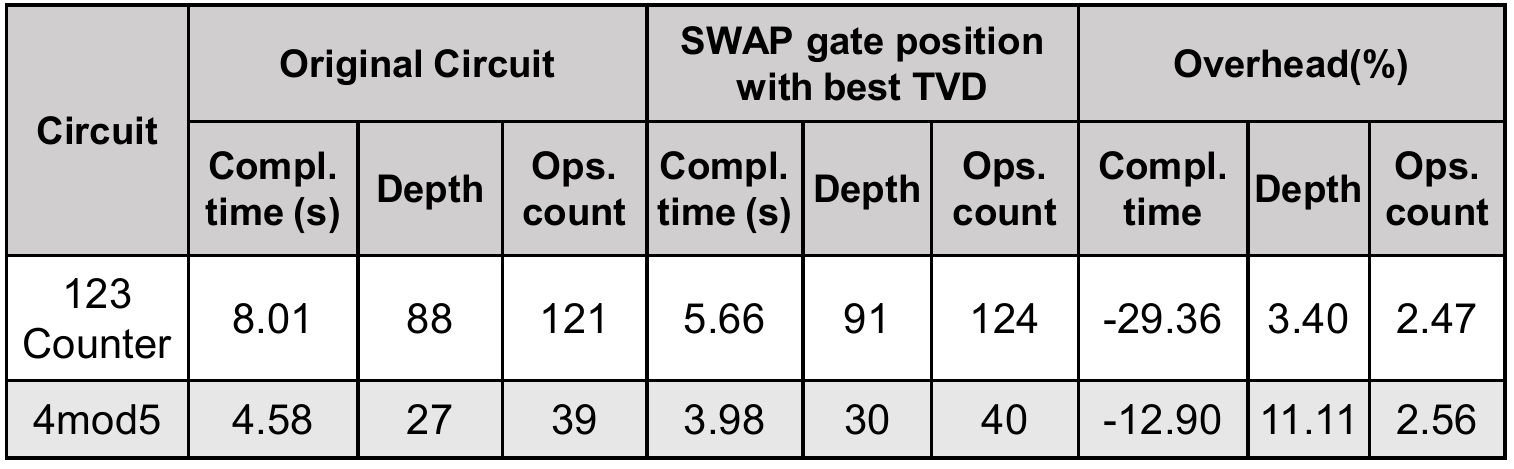}
    \captionof{table}{Design overhead for obfuscation in terms of circuit depth, number of basis operations, and compilation time for two benchmark circuits with the Qiskit compiler backend.}
    \label{fig:oh}
\end{figure}
{\bf{Overhead analysis:}} The additional dummy SWAP gates and circuit barriers may affect the compiler performance. To investigate the potential impacts, we compile two benchmarks - 123\_counter, and 4mod5, with (a single dummy SWAP) and without obfuscation using the Qiskit compiler \cite{cross2018ibm} and compare the circuits depth, gate-count, and compilation time as shown in Table \ref{fig:oh}. The circuit depth increased by 3.4\% and 11.11\% each as expected. Similarly, gate-count increased by 2.46\% and 2.56\%. However, the compilation time decreased by 29.36\% and 12.9\%. 
The added barriers prevented aggressive circuit optimization which translated to the smaller compilation time. Note that, we can improve the strength of obfuscation by adding more dummy gates into the circuit at the expense of higher circuit depth and gate-count upon compilation.

\section{Discussions} \label{disc}

\subsubsection{Consideration for multiple dummy gates}
Addition of multiple dummy SWAP gates can be considered to obtain higher TVD and to increase the adversarial effort further. However, the overhead will also be higher. This aspect can be investigated in future research.
\subsubsection{Impact of dummy gates on coherence} We insert dummy SWAP gates only in the compilation phase. Before executing the circuit on the real hardware, the dummy gates will be removed.
The device executable version of the circuit will not contain any additional gates. Therefore, the proposed approach does not affect coherence during execution but will secure the circuit during compilation on third-party compilers.
\subsubsection{Consideration for non-arithmetic circuit} Non-arithmetic quantum circuits can be studied in future research to identify other metrics that can be used for obfuscation. These circuits will provide a better understanding since measurement in X-basis and Hadamard basis are not commonly covered in arithmetic circuits.
\subsubsection{Recognition and removal of added extra logic} After compilation, the designer needs to identify and remove the extra dummy gates to retrieve the original circuit functionality. This could pose a challenge since the dummy SWAP gate could be, (i) optimized with other gates in the circuit; (ii) removed during compilation if a reverse SWAP is required to meet the coupling constraint of the hardware; and, (iii) mixed up with other gates and could be hard to identify due to change in circuit depth and other add SWAP gates. 
As an initial solution, we employed the barriers to enclose the dummy SWAP gate. This prevents the compiler from optimizing the SWAP gate with other gates. Although the added barrier could lead to slight degradation in optimization, it provides an easy mechanism to the designer to identify and remove the dummy SWAP gate post compilation. Note that the added barrier may provide clue to the adversary. One can obfuscate such clues by adding dummy barriers in the design.   
\subsubsection{Usage of noisy quantum simulator} Real quantum hardware inherently provides probabilistic outputs due to various sources of known and unknown errors that are an active area of research such as, cross-talk. This can present a challenge for analysis and derivation of clear intuition for the development of a metric for guided SWAP insertion. For example, the crosstalk error can overshadow the benefits of a certain metric. To minimize the errors in analysis, we employ noisy quantum simulator with well characterized noise models such as, gate error and decoherence for our analysis.

\subsubsection{Adversarial reverse engineering effort} As mentioned earlier, oracle-based attack is not possible in quantum domain. Adversary could be unaware of the obfuscation in the circuit and as a result, would get incorrect netlist. Even if the adversary is aware that the circuit has been obfuscated, the reverse engineering effort of locating the dummy SWAP gates will be extremely hard due to lack of oracle model.
The adversary can try removing gates from the obfuscated circuit to retrieve the original circuit. In our example circuits, we inserted only one dummy gate. For such case, the search complexity is $O(n)$ where $n$ is the number of gates in the obfuscated circuit. This linear time complexity however, there is no way to validate the guess. Therefore, adversary will end up with $n$ possible circuit out of which one could correct. 
\section{Conclusion} \label{conc}

Future quantum computing workload will rely on untrusted $3^{rd}$ party for optimization of large scale quantum circuit. This can expose them to threats such as, counterfeiting and reverse engineering. We presented quantum circuit obfuscation by inserting two qubit SWAP gates and 
quantified the 
effectiveness using TVD based metric. We also presented an automated procedure to select the best position for dummy SWAP gate insertion which maximizes the TVD without requiring time intensive simulation of quantum circuit. The proposed metric achieved approximately $5-7\%$ improvement than average TVD, and approximately $12-13\%$ closer to the best TVD at an overhead of less than $5\%$ for the number of gates post compilation.

\bibliographystyle{IEEEtran}
\bibliography{IEEEabrv,ref}
\EOD
\end{document}